\documentclass[journal]{IEEEtran}
\usepackage[noadjust]{cite}
\usepackage{graphicx}
\usepackage{subfigure}
\usepackage{amsmath,lipsum}
\usepackage{float}
\usepackage{color, soul}
\usepackage{cite}
\usepackage{multicol}
\usepackage{multirow}

\usepackage{amsmath}
\usepackage{tabularx}

\usepackage{hyperref}
\hypersetup{
    pdftitle={Carbon and Reliability-Aware Computing for Heterogeneous Data Centers},    
    pdfauthor={},     
    pdfsubject={},   
    pdfcreator={},   
    pdfproducer={},  
    pdfkeywords={}, 
}

\begin{document}


\title{Carbon and Reliability-Aware Computing for Heterogeneous Data Centers}

\author{
    Yichao Zhang, \IEEEmembership{Member, IEEE},
    Yubo Song, \IEEEmembership{Member, IEEE}, and
    Subham Sahoo, \IEEEmembership{Senior Member, IEEE}
    \thanks{All authors are with the Department of Energy, Aalborg University, 9220 Aalborg, Denmark, with e-mail addresses: \{\texttt{yzha, yuboso, sssa}\}@energy.aau.dk.}
}

\maketitle

\begin{abstract}
The rapid expansion of data centers (DCs) has intensified energy and carbon footprint, incurring a massive environmental computing cost. While carbon-aware workload migration strategies have been examined, existing approaches often overlook reliability metrics such as server lifetime degradation, and quality-of-service (QoS) that substantially affects both carbon and operational efficiency of DCs. Hence, this paper proposes a comprehensive optimization framework for spatio-temporal workload migration across distributed DCs that jointly minimizes operational and embodied carbon emissions while complying with service-level agreements (SLA). A key contribution is the development of an embodied carbon emission model based on servers’ expected lifetime analysis, which explicitly considers server heterogeneity resulting from aging and utilization conditions. These issues are accommodated using new server dispatch strategies, and backup resource allocation model, accounting hardware, software and workload-induced failure. The overall model is formulated as a mixed-integer optimization problem with multiple linearization techniques to ensure computational tractability. Numerical case studies demonstrate that the proposed method reduces total carbon emissions by up to 21\%, offering a pragmatic approach to sustainable DC operations.
  
\end{abstract}
\begin{IEEEkeywords}
Carbon-aware computing, Server reliability, Backup resource allocation, Server heterogeneity.
\end{IEEEkeywords}

\section{Introduction}

\IEEEPARstart{T}{he} rapid proliferation of data centers (DCs) has become the cornerstone of the modern digital economy, supporting a myriad of services from cryptocurrencies to artificial intelligence (AI). However, DCs have recently proven to be one of the most energy-intensive infrastructures. In 2022, global DCs consumed approximately 460 TWhs of electricity, accounting for nearly 2\% of the global electricity demand \cite{IEA2024electricity}. With the increasing adoption of large language models (LLMs), the electricity consumption of data centers is projected to increase by 165\% due to AI workloads by 2030, compared to the power demand in 2023 \cite{GoldmanSachs2024, IEA2024}. Since a large share of electricity is still generated from carbon-intensive energy resources, the increased power demand from DCs can lead to significant carbon emissions with an instinctively high environmental cost. 

Since \textit{carbon intensity} (the amount of carbon emissions per unit electricity generation) vary significantly across spatio-temporal scales due to renewable energy generation variability \cite{8333479,10733074} and the amount of computing workloads of geo-distributed DCs \cite{Chen2020, Zeng2024}, an emerging solution addresses this issue by aligning workloads based on the carbon intensity from electrical sources. This leads to novel \textit{carbon-aware} DC operation strategies, including temporal \cite{9770383}, spatial \cite{8086181}, or spatio-temporal \cite{9627522, 10518095, 10591261} migration of workloads. The authors in \cite{9770383} propose a representative carbon-aware mechanism that employs virtual capacity curves to facilitate temporal workload shifting based on the predicted carbon intensity, such that the critical workloads are prioritized during lower emission periods. For spatial workload migration, authors in \cite{8086181} propose a financial incentive mechanism for geo-distributed DCs to encourage more workloads execution during lower carbon-intense period, especially when the workloads are time-sensitive. Furthermore, spatio-temporal migration strategies \cite{9627522, 10518095, 10591261} jointly optimizes workload execution time and location to reduce carbon emissions by aligning them based on renewable generation profile.

Although the aforementioned \textit{carbon-aware} workload management and computing approaches are well-positioned from a sustainable view-point, the operational durability of DCs is in fact overlooked -- which is a critical dimension that evaluates the lifespan management of servers and their failure risk mitigation. In particular, the following constraints can be subsequently forewarned for a comprehensive life cycle assessment of DCs:

\begin{enumerate}
    \item \textit{\textbf{Embodied CO$_2$ emissions:}} Although \cite{8361052} suggests that the proportion of replacement costs in overall expenditure can be decreased from 80\% to 20\% by including regular maintenance of DCs, the \textit{embodied} carbon emissions, generated during server manufacturing, transportation, and end-of-life recycling \cite{Dirty, Busa2019, Wu2021SustainableAE}, amounts upto 30\%-50\% of the total lifetime emissions. This is exceptionally significant, given the iteration and/or upgrade of AI servers in recent years \cite{Bodner2025}. On a contrary, the induced embodied carbon emission cost is not among the considerations of \cite{9627522, 10518095, 10591261}, which then yields a sub-optimal schedule. Inspired by \cite{8468102} and \cite{li2025ecoserve}, this work recognizes the degradation of servers as a significant and pragmatic concern aggregating alongside the spatio-temporal workload scheduling, which has the potential to further reduce the carbon footprint of DC operations.
    
    \item \textit{\textbf{Failure risks of DC servers:}} Neglecting failure risks will lead to a decline in quality of service (QoS), or violations of service level agreements (SLA), thereon an increased cost and service reputation. An operational record from Google \cite{9248891} showcases that over 40\% of servers fails at least once during 29 days, where the cost of ensuing outage can reach as much as \$5,600/min and may exceed \$1 million. This necessitates the focus on DC outages and proactive failure risk management.  
    To this end, \cite{9926075, 8422840} have inspected the role of redundancy (backup servers) to minimize the service interruption periods, while it can be more profitable to dynamically trade-off between carbon cost and operation durability. In this paper, we tend to incorporate both the backup resources planning \cite{9926075, 8422840} and lifetime degradation modeling of DCs into the spatio-temporal workload scheduling, to better mitigate the SLA violations and guarantee their legitimacy.
    
    \item \textit{\textbf{Server heterogeneity:}} Optimal accuracy may be compromised by the heterogeneity among servers. In existing research \cite{9627522, 10518095, 10591261}, a general assumption is the uniformity of the servers, where model uncertainties are not properly justified. Although it benefits by a simpler computation task, their robustness may be undermined. Meanwhile, it has been explicated that servers in a DC exhibit distinct aging rates \cite{Vishwanath2010} and performance qualities \cite{ReAD2022,1010076,wang2023giving}, where older servers are typically more prone to wear-out failures and should be carefully dispatched. Hence, it is vital to identify server provisioning in accordance with their aging performance differences in spatio-temporal workload scheduling and improve their operational durability and efficiency.
\end{enumerate}

In this context, we deep dive into this topic and fill in the aforementioned research gaps. We examine the impact of lifetime, failure risk, and heterogeneity of servers on the spatio-temporal workload migration strategy. The major contributions of this paper include:
\begin{enumerate}
    \item We propose a chance-constrained backup resource allocation model to address workload and server (hardware, software) based failures in compliance with SLA.
    \item We unravel the decisive factors for more liable DC operations, including utilization rates, uncertainties in aging performance distribution, and failure probability, that accumulates over time. The server provisioning strategy is then derived to ensure precise workload allocation considering server heterogeneity.
    \item We formalize a management framework for DCs that can balance carbon footprint and DC operational durability, and provide comprehensive insights in maneuvering the optimal migration and operation strategies. 
\end{enumerate}
As compared to the existing approaches, the solutions derived from the proposed framework are pragmatic and provides a thorough sustainable assessment for operators by incorporating the lifetime analysis, backup resource reservation, and server heterogeneity. 

\begin{figure}[!h]
    \centering
    \includegraphics[width=\linewidth]{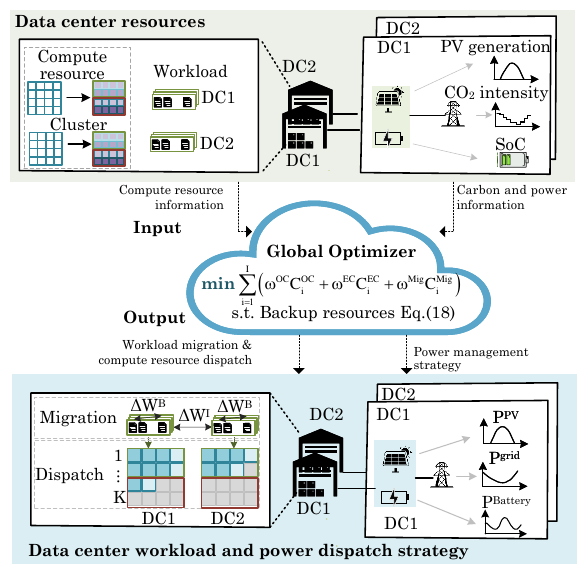}
    \vspace{-9pt}
    \caption{Proposed framework for distributed DCs operation -- accounting failure and server heterogeneity. $\mathrm{P^{PV}}$,  $\mathrm{P^{Grid}}$, and $\mathrm{P^{Battery}}$ represent the power from photovoltaic (PV), power grid, and battery, which satisfy the power balance. Additionally, $\mathrm{\Delta W^{B}}$ and  $\mathrm{\Delta W^{I}}$ represent the migrated batch and interactive workload strategy, governed by \eqref{eq:WI1} to \eqref{eq:BI1}. K server clusters are dispatched to accommodate the workload after migration according to their heterogeneity.}
    \label{Framework}
    \vspace{-9pt}
\end{figure}

\section{Proposed DC Operation Framework}

Fig. \ref{Framework} illustrates the proposed framework of the DCs operation strategy. Two types of carbon emission are considered here:

\begin{itemize}
    \item \textbf{Operating carbon emissions $\mathrm{C^{OC}}$}: This concerns the power consumption from the grid and its associated carbon emission intensity.
    \item \textbf{Embodied carbon emissions $\mathrm{C^{EC}}$}. This is incurred by server and component replacement during manufacturing. 
\end{itemize}

As demonstrated in Fig. \ref{Framework}, the global optimizer first aggregates information from distributed DCs, including computational resources (server clusters and workloads), carbon emission intensity, and local power generation profiles (PV and BESS). Subsequently, it co-optimizes $\mathrm{C^{OC}}$ and $\mathrm{C^{EC}}$ using an integrated framework by coordinating workload migration strategies, heterogeneous computing resource dispatching, and power management schemes. In this process, a backup resource deployment strategy is also incorporated to respond to failure events.

Existing studies that assume server homogeneity may lead to stochastic workload allocation among servers. However, this work explicitly considers server heterogeneity, which arises due to two critical segments:
 \begin{enumerate}
    \item \textbf{Embodied carbon divergence}: Operation histories vary significantly across servers, which leads to a divergent degradation process, replacement cycle, and manufacturing-related embodied carbon footprint. These differences necessitate a compute resource dispatch strategy considering heterogeneity to prolong servers' lifetime while minimizing daily embodied carbon emissions.
    \item  \textbf{Dynamic failure probability}: Server failure probability escalates with cumulative operating time, requiring proactive deployment of redundant backup resources to ensure reliability.
\end{enumerate}

To address these challenges, our proposed compute resource dispatch strategy incorporates:
\begin{enumerate}
    \item \textbf{Heterogeneity-aware clustering}: Servers are clustered based on their history of operational features/patterns (e.g., operating time, failure history).
    \item  \textbf{Failure-adaptive redundancy}: Backup resources are allocated proportionally to cluster-specific failure probabilities.
\end{enumerate}

\begin{figure}[!h]
    \centering
    \includegraphics[width=0.95\linewidth]{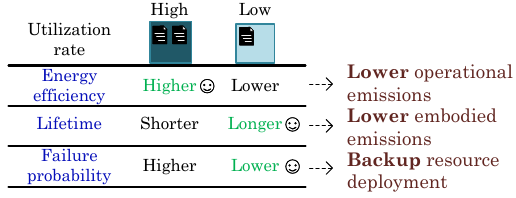}
    \vspace{-9pt}
    \caption{Impact of server utilization rate on energy efficiency, lifetime, and failure probability and its consequence on operation economy and emissions.}
    \label{utilization}
\end{figure}

Notably, the embodied and operation carbon footprint, along with backup resource deployment in the proposed framework, are linked to the server utilization rate $\mathrm{u}$. As illustrated in Fig.~\ref{utilization}, higher utilization improves the DCs' energy efficiency by reducing the number of active servers required to handle workloads. This lowers overall energy demand and subsequently decreases operational emissions. However, increased utilization also accelerates hardware wear and tear, leading to shorter server lifespans and a higher failure probability, which in turn necessitates more frequent maintenance or replacements. On the other hand, server lifespan is directly linked to embodied carbon emissions, as shorter lifespans amplify the environmental impact of hardware manufacturing and replacement. Additionally, higher failure probability increases the need for backup resource deployment to ensure service reliability. Hence, this paper will also investigate the combinatorial outcome of the abovementioned factors by analyzing the impact of server utilization rate $\mathrm{u}$ on DC operation.
\vspace{-10pt}

\vspace{-20pt}

\subsection{Optimization function}

The operation cost of a total number of $\mathrm{I}$ DCs will be optimized at the same time using: 
\begin{equation}
    \min \mathrm{\sum\nolimits_{i=1}^{I} {\left( {\omega ^{OC} C_i^{OC} + \omega ^{EC} C_i^{EC} +\omega ^{Mig}  C_i^{Mig} } \right)}}
    \label{eq:obj}
\end{equation}
where, $\mathrm{C_i^{Mig}}$ is the workload migration cost, and $\mathrm{i}$ is the DC index. $\mathrm{\omega ^{OC}}$, $\mathrm{\omega ^{EC}}$, and $\mathrm{\omega ^{Mig}}$ are weighting coefficients that reflect the relative importance of the three items.
\subsubsection{Operation cost $\mathrm{C^{OC}}$} To minimize the operational carbon cost of DCs, we require the grid power consumption $\mathrm{P_{i}^{grid}(t)}$ and regional carbon intensity $\mathrm{CI^{OC}_{i}(t)}$, monetized through the coefficient of carbon emission cost  $\mathrm{\alpha^{C}}$ ($\mathrm{\$/kgCO_2}$)  using:
\begin{equation}
    \mathrm{C_i^{OC} = \sum\nolimits_{t=1}^{24} {\alpha ^{C}CI^{OC}_{i}(t)P_{i}^{grid}(t)\Delta t}} 
    \label{eq:Carcost}
\end{equation}
where, $\mathrm{\Delta t}$ is the time interval of the operating strategy, which is 1 hour.
 
\subsubsection{Embodied carbon emission} 

The daily embodied carbon emissions $\mathrm{C_{i}^{EC}}$ of $\mathrm{i^{th}}$ DC is the sum of the embodied carbon emission from $\mathrm{K}$ server groups, given by:
\begin{equation}
    \mathrm{C_{i}^{EC} = \sum\nolimits_{k=1}^{K} {C_{k,i}^{EC}}}
    \label{eq:EC1}
\end{equation}
To estimate the daily embodied carbon emission $\mathrm{C_{k, i}^{EC}}$ of the server group $\mathrm{k}$ in $\mathrm{i^{th}}$ DC, a model based on the entire lifetime of the servers or its component is proposed: 
\begin{equation}
    \mathrm{C_{k,i}^{EC} = {\alpha ^C}CI_{k,i}^{EC}{N_{k,i}}\Delta {t^{Eq}_{k,i}}} 
    \label{eq:EC2}
\end{equation}
where, $\mathrm{N_{k,i}}$ is the number of servers. $\mathrm{\Delta {t^{Eq}_{k,i}}}$ is measured in units of day, corresponding to the equivalent daily operating time of server cluster $\mathrm{k^{th}}$.
The daily embodied carbon intensity $\mathrm{CI_{k,i}^{EC}}$ is calculated using \eqref{eq:EC3}, where the manufacturing carbon footprint $\mathrm{C_{k,i}^{M}}$ of $\mathrm{N_{k,i}}$ servers is distributed into its expected calendar lifetime $\mathrm{T_{k,i}^{Can}}$.
\begin{equation}
    \mathrm{CI_{k,i}^{EC} = \frac{{C_{k,i}^{M} \cdot N_{k,i}}}{T_{k,i}^{Can}}}
    \label{eq:EC3}
\end{equation}

\begin{figure}[t]
    \centering
    \includegraphics[width=0.9\linewidth]{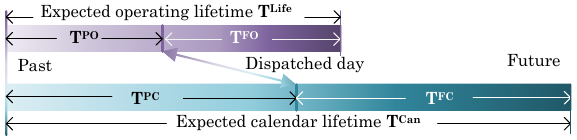}
    \vspace{-9pt}
    \caption{Relationship between expected calendar lifetime and expected operating lifetime. Expected server’s calendar lifetime $T^{\mathrm{T^{Can}}}$ consists of past calendar operating time $T^{\mathrm{T^{PC}}}$ and future calendar operating time $T^{\mathrm{T^{FC}}}$.  $T^{\mathrm{T^{FC}}}$ is estimated by the expected future operating lifetime $T^{\mathrm{T^{FO}}}$ and operation during the dispatched day.}
    \label{lifetime}
    \vspace{-9pt}
\end{figure}

As shown in Fig. \ref {lifetime} and defined in \eqref{eq:EC4}, the expected calendar lifetime $\mathrm{T_{k,i}^{Can}}$ of server group $\mathrm{k}$ should be calculated based on its past calendar operating time $\mathrm{T_{k,i}^{PC}}$ and its expected future calendar operating time $\mathrm{T_{k,i}^{FC}}$. It is worth notifying that the past calendar operating time can be easily obtained by checking the records of server or component purchases.
  
\begin{equation}
    \begin{split}
    \mathrm{T_{k,i}^{Can}} &= \mathrm{\left( {T_{k,i}^{PC} + T_{k,i}^{FC}} \right) {N_{k,i}}}\\
    &= \mathrm{\left( {T_{k,i}^{PC} + T_{k,i}^{RO}\frac{{{T^{day}}}}{{\Delta {t^{Eq}}}}} \right) {N_{k,i}}}
    \end{split}
    \label{eq:EC4}
\end{equation}
\begin{equation}
    \mathrm{T_{k,i}^{FO} = T_{k,i}^{Life} - T_{k,i}^{PO}}
    \label{eq:EC5}
\end{equation}
where, $\mathrm{T_{k,i}^{FO}}$ is the remaining operation time. By dividing the daily consumed operating time $\mathrm{\Delta t_{k,i}^{Eq}}$ by $\mathrm{T_{\text{day}}}$, the operating time consumed speed can be estimated, further deducing the expected future calendar time based on the remaining operating time $\mathrm{T_{k,i}^{on}}$. The daily equivalent operating time $\mathrm{\Delta t_{k,i}^{Eq}}$ is given by:
\begin{equation}
    \begin{split}
    \mathrm{\Delta {t^{Eq}}} &= \mathrm{\frac{{\sum\nolimits_{t=1}^{24} {N_{k,i}^A(t)} \Delta t}}{{\sum\nolimits_{t=1}^{24} {{N_{k,i}(t)}} \Delta t}}{T^{day}}}=\frac{\sum\nolimits_{t=1}^{24} {N_{k,i}^A(t)}}{24{N_{k,i}}}
    \end{split}
    \label{eq:EC6}
\end{equation}
where, $\mathrm{N_{k,i}^{A}(t)}$ is the number of dispatched servers in the group $\mathrm{k}$ of DC $\mathrm{i}$ at time $\mathrm{t}$ and $\mathrm{N_{k,i}(t)}$ denotes the number of total servers in group $\mathrm{k}$ of DC $\mathrm{i}$ at time $\mathrm{t}$.  On top, $\mathrm{T^{day}}$ denotes the length of the dispatched day, which is 1 day. $\mathrm{\Delta t_{k,i}^{Eq}}$ take 1 day as unit.

Using (3)-(8), the daily embodied carbon emission $\mathrm{C_{i}^{EC}}$ can be given by:
\begin{equation}
    \mathrm{C_i^{EC} = \sum\nolimits_{k=1}^{K} {\frac{{C_{k,i}^{M} {{\left( {\sum\nolimits_{t=1}^{24} \cfrac{N_{k,i}^{A}(t)}{24} } \right)}^2}}}{{T_{k,i}^{PC} \sum\nolimits_{t=1}^{24} {\cfrac{N_{k,i}^{A}(t)}{24}}  + T_{k,i}^{FO} {N_{k,i}}}}}} 
    \label{eq:EC6}
\end{equation}

\subsubsection{Migration cost} This caters to the cost function associated with transferring workloads between DCs, given by:
\begin{equation}
    \mathrm{C_i^{Mig} =  {{\alpha ^{Mig}}{{\left| {\Delta W_{i,t}^I} \right|}}}} ,{\kern 1pt} {\kern 1pt} \ \mathrm{\forall i \in I}
    \label{eq:migracost}
\end{equation}
where, $\mathrm{\alpha ^{Mig}}$ is the coefficient of migration cost with unit of  $\mathrm{\$/requests/s}$.

\subsection{Workload Migration Model}

\subsubsection{Interactive Workload (IW) Modeling}
IWs refer to computing tasks or applications that require real-time interaction and responses, such as real-time payments and online services, which are \textit{delay-sensitive}. As a result, such workloads have spatial flexibility, which can be routed to another DC immediately as per the availability of computing resources. The migration model of interactive workload between different DCs can be modeled using:
\begin{equation}
    \mathrm{W_{i}^I(t) = W_{i}^{I,ini}(t) - \Delta W_{i}^I(t)	\ni   {W_{i}^I(t) \geq 0}}
    \label{eq:WI1}
\end{equation}
\begin{equation}
    \mathrm{\Delta W_{i}^I(t) = \sum\nolimits_{j \in I,j \ne i} {\Delta W_{i \to j}^I(t)}} \
    \label{eq:WI2}
\end{equation}
where, $\mathrm{W_{i}^{I,ini}(t)}$ and $\mathrm{W_{i}^I(t)} $ denote the interactive workload before and after migration at time $\mathrm{t}$ in DC $\mathrm{i}$, respectively. Furthermore, $\mathrm{\Delta W_{i}^I(t)}$ in (12) denotes the total received or migrated workload at time $\mathrm{t}$ in DC $\mathrm{i}$. It should be noted that $\mathrm{\Delta W_{i}^I(t)}$ is positive, when the workload is offloaded/moved out. To ensure all interactive workloads can be processed immediately, all migrated workloads should be processed such that $\mathrm{\sum\nolimits_{i=1}^{I} {\Delta W_{i}^I(t)} = 0}$. Furthermore, the amount of transferred workload is limited within the upper limit of the transmission fiber $\mathrm{\Delta W_{i \to j}^{I,ub}}$, given by:
\begin{eqnarray}
    \mathrm{| \Delta W_{i \to j}^{I}(t)|\le\Delta W_{i \to j}^{I,ub}} 
    \label{eq:WI4}
\end{eqnarray} 		
\subsubsection{Batch Workload (BW) Modeling}
BWs refer to the \textit{flexible} tasks that do not need to be addressed immediately and are sufficient to be completed before the deadline, such as data analysis and model training. Although this facilitates temporal flexibility for batch workloads while lacking spatial adjustment, they heavily rely on large-scale databases, which will incur significant migration costs. 

Based on Google's batch workload migration strategy \cite{9770383}, the batch workload transfer model is given by:
\begin{equation}
    \mathrm{W_{i}^B(t) = \frac{ \sum\limits_{t=1}^{24}{W_{i}^{B,ini}(t)}} {24} + \Delta W_{i}^B(t)} \ni {\small\begin{cases} \mathrm{W_{i}^B(t) \geq 0} \\ \mathrm{\sum\limits_{t=1}^{24} {\Delta W_{i}^B(t)}  = 0} \end{cases}}
    \label{eq:BI1}
\end{equation}
where, $\mathrm{W_{i}^{B,ini}(t)}$, $\mathrm{W_{i}^B(t)} $ and $\mathrm{\Delta W_{i}^B(t)}$ denote the batch workload before, after migration and the amount of transferred batch workload at time $\mathrm{t}$ in DC $\mathrm{i}$, respectively.

\subsection{Computing Resource Dispatch Strategy}  

Considering server heterogeneity and failure events, it is vital to account for a computing resource dispatch strategy and optimize workload allocation into available servers. To be specific, the server cluster and the number of primary and backup servers in each cluster will be optimized for dispatching decisions.

The workload accommodated by $\mathrm{k}$ server cluster are allocated in a dynamic manner using:
\begin{subequations}
    \begin{equation}
        \mathrm{W_{i}^I(t) = \sum\nolimits_{k=1}^{K} {W_{k,i}^I(t)}} 
   \end{equation}	
    \begin{equation}
        \mathrm{W_{i}^B(t) = \sum\nolimits_{k=1}^{K} {W_{k,i}^B(t)} }  
    \end{equation}	
    \label{eq:N1}
\end{subequations}
where, $\mathrm{W_{k,i}^I(t)}$ and $\mathrm{W_{k,i}^B(t)}$ denote the accommodated interactive and batch workloads in DC $\mathrm{i}$. It should be noted that the primary servers in cluster $\mathrm{k}$ should be sufficient to accommodate the total assigned workloads, as defined in: 
\begin{equation}
    \mathrm{0 \le [W_{k,i}^I(t) + W_{k,i}^B(t)] \le N_{k,i}^P(t) \cdot {S_{k,i}^{rate}} \cdot {u_{k,i}(t)}}
    \label{eq:N2}
\end{equation}
where, $\mathrm{N_{k,i}^P(t)}$, $\mathrm{S_{k,i}^{rate}}$, $\mathrm{u_{k,i}}$ denote the number of dispatched primary servers, the maximum speed of servers to deal with workloads, and the server utilization rate of the server cluster $\mathrm{k}$ in DC $\mathrm{i}$, respectively.

\begin{figure}[t]
    \centering
    \includegraphics[width=0.9\linewidth]{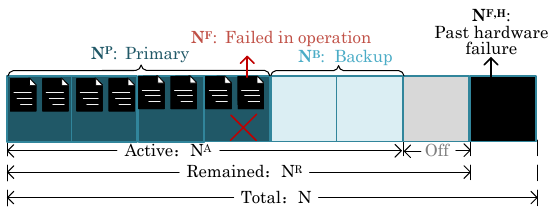}
    \vspace{-9pt}
    \caption{Number of servers in different statuses. A server cluster includes $\mathrm{N}$ servers during the dispatched day. At time $\mathrm{t}$, there are $\mathrm{N^{F,H}}$ accumulated failed servers due to hardware failure, thus the remaining available servers are $\mathrm{N^{R}}$. $\mathrm{N_{A}}$ servers are in active status to deal with workload or respond to failure events during the operation.}
    \label{Server relationship}
    \vspace{-9pt}
 \end{figure}

During the operating process of DCs, the dispatched servers should satisfy the relationship defined in Fig. \ref{Server relationship}. In a server cluster with $\mathrm{N_{k,i}}$ servers, except for the deployment of primary servers, backup resources $\mathrm{N_{k,i}^B(t)}$ should be reserved considering the failure events during server operation. To ensure sufficient backup resources under uncertain failure events, a chance-constrained programming is proposed, as follows: 
\begin{equation}
    \mathrm{Pr \left( N_{k,i}^F(t) \leq N_{k,i}^B(t) \right) \geq p^{thr}}
    \label{eq:N3}
\end{equation}
where, $\mathrm{N_{k,i}^F(t)}$ are the failed resources at time $\mathrm{t}$. More details on quantifying $\mathrm{N_{k,i}^F(t)}$ can be found in Section III.  $\mathrm{p^{thr}}$ is the target threshold probability. This constraint ensures that with at least $\mathrm{p^{thr}}$ confidence, the reserved backup servers are sufficient to cover the random failure events.

Hence, the servers with an \textit{active} status $\mathrm{N_{k,i}^A(t)}$ that can be dispatched are given by:
\begin{subequations}
    \begin{alignat}{2}
        \mathrm{N_{k,i}^P(t) + N_{k,i}^B(t) = N_{k,i}^A(t)}\\
        \mathrm{N_{k,i}^P(t) \ge 0, N_{k,i}^B(t)\ge 0}
    \end{alignat}
    \label{eq:N4}
\end{subequations}
Given that there will be failure events, servers that can be fully exploited will decrease with time. Software failures can typically be resolved by restarting the affected servers, leading to only a temporary reduction in the current hour. However, since hardware failures require a complex and time-consuming repair process, affected servers will not be available for the remainder of the dispatch day. Consequently, it results in a reduction in the remaining available servers $\mathrm{N_{k,i}^R}$, as follows:
\begin{equation}
    \mathrm{N_{k,i}^R(t) = N_{k,i} - \sum\nolimits_{\tau  = 1}^t {N_{k,i}^{F,H}(\tau)} }
    \label{eq:N5}
\end{equation}
where, $\mathrm{N_{k,i}^R(t)}$, $\mathrm{N_{k,i}^{F,H}(\tau)}$ denote the remaining available servers in cluster $\mathrm{k}$, and the number of failed servers due to hardware failure at time $\mathrm{\tau}$, respectively. More details on determining $\mathrm{N_{k,i}^{F,H}(\tau)}$ can be found in Section III.

Therefore, the dispatched active servers $\mathrm{N_{k,i}^A(t)}$ should not be above the remaining available servers $\mathrm{N_{k,i}^R(t)}$:
\begin{equation}
    \mathrm{0 \le N_{k,i}^A(t) \le N_{k,i}^R(t)}
    \label{eq:N6}
\end{equation}
 
\subsection{Service level agreement (SLA) constraint}

The execution time of each workload must comply with strict SLA constraints, which are quantified by the execution delay time $ \mathrm{t^{delay}_{k, i}(t)} $. This delay time is influenced by the server resource utilization rate $\mathrm{u_{k,i}(t)}$, as described in:
\begin{equation}
    \mathrm{t^{delay}_{k,i}(t)=\frac{1}{{(1 - u_{k,i}(t))S^{rate}}} \le t^{delay,ub}}
    \label{eq:delay}
\end{equation}
To ensure compliance with SLA constraints, the utilization rate $ \mathrm{u_{k,i}(t)}$ must always be limited within upper boundary $ \mathrm{u^{\text{ub}}_{k,i}} $ at any given time $\mathrm{t}$.
\begin{equation}
    \mathrm{u_{k,i}(t) \le u^{\text{ub}}_{k,i}}
    \label{eq:utilization}
\end{equation}

\subsection{Power consumption model of DCs}

The power consumption of DC $\mathrm{i}$ can be estimated based on the number of active servers, its utilization rate, and the power usage efficiency $\mathrm{PUE_i}$ \cite{10518095} using:
\begin{equation}
\begin{split}
    \mathrm{P_{i}^{DC}(t)} &= \mathrm{\sum\nolimits_{k=1}^{K} \Big[{(P_{k,i}^{idle} + (PU{E_i} - 1)P_{k,i}^{peak})N_{k,i}^A(t)}} \\
    & \mathrm{+ {u_{k,i}(t)}\left( {N_{k,i}^A(t) - N_{k,i}^F(t)} \right)(P_{k,i}^{peak} - P_{k,i}^{idle})\Big]}
\end{split}
\label{eq:PC}
\end{equation}
where, $\mathrm{P_{k}^{idle}(t)}$ and $\mathrm{P_{k,i}^{peak}}$ are the idle and peak power of a server, respectively. Besides, each DC must satisfy the power balance constraints, which can be found in detail in \cite{10518095}.

\section{Reliability and Redundancy Model of Servers}

This section focuses on a deployment strategy for computing backup resources, which comprehensively considers the loss of computing resources due to hardware failure, software failure, and schedule failure.

\subsection{Principle of Backup Resources Strategy}

To ensure the reliability of data centers (DCs), backup servers must be deployed alongside primary servers. Under normal conditions, primary servers handle workloads, process requests, and maintain system performance. However, in the event of a primary server failure or failed/evicted scheduled workloads, backup servers, which serve as redundant units, immediately take over these workloads to maintain uninterrupted operation.

Both primary and backup servers operate in \textit{active} status to enable seamless operation, accommodate additional workload, and maintain primary server performance. However, even in \textit{idle} status, backup servers consume approximately half of the peak power, increasing carbon emissions. Hence, a accurate determination of the number of backup servers required is essential to balance energy efficiency and service availability.

\subsection{Reliability Modeling of Servers}

To quantify the required backup resources, this subsection firstly focuses on quantifying the reliability of servers. As per the statistics in \cite{4775906}, primary reasons that lead to server failure in DCs can be categorized into: hardware, software, and humans. Among them, server failures caused by hardware and software errors account for upto 70\%. Hence, we narrow down our focus on the design of failure probability models originating out of hardware and software failure to quantify redundancy needs.

\subsubsection{Hardware reliability}
Different from the current server reliability models that neglect servers' past operating time with an assumption of a constant failure rate, this paper builds on the theory that hardware failure of servers primarily stemming from the cumulative wear and tear of critical components. Hence, Weibull distribution \cite{4775906} is employed to formalize hardware reliability based on a time-dependent degradation process, given by: 
\begin{equation}
    \mathrm{{p^H} = \exp\left[{-\left( \frac{t^{on}}{\psi} \right)}^\beta \right]}
    \label{eq:Weib}
\end{equation}
where, $\mathrm{\beta}$ is the shape parameter of the Weibull distribution. $\mathrm{\psi}$ is the scale parameter of the Weibull distribution at base utilization rate $\mathrm{u^{base}}$, representing the characteristic lifetime of the hardware $\mathrm{\psi}$ used with a base utilization rate $\mathrm{u^{base}}$. Basically, (\ref{eq:Weib}) defines the cumulative probability $\mathrm{p^{H}}$ for a server that can operate reliably (without any failure) until time $\mathrm{t^{on}}$. On this basis, the probability of servers surviving throughout the dispatched day ($\mathrm{t^d+T^{d,eq}}$) should be modeled as a conditional probability function, as formulated below:
\begin{equation}
    \begin{split}
        \mathrm{{p^H}}&\mathrm{\left( {\left. {{t^{on}} \geq {t^d} + T^{d,eq}} \right|{t^{on}} \geq {t^d}} \right)}\\
        &=\mathrm{ \frac{{{p^H}\left( {{t^{on}} \geq {t^d} + T^{d,eq}} \right)}}{{{p^H}\left( {{t^{on}} \geq {t^d}} \right)}}}\\
        &= \mathrm{\exp{\left[{\left(\frac{t^d}{\psi} \right)^\beta} - {\left(\frac{t^d+T^{d,eq}}{\psi} \right)^\beta} \right]}}
    \end{split}
    \label{eq:HardCondition}
\end{equation}
where, $\mathrm{T^{d,eq}}$ represents the equivalent consumed lifetime of a server when operating for one day at utilization $\mathrm{u}$, normalized to the baseline utilization rate $\mathrm{u^{base}}$. The corresponding mathematical formulation is given by:
\begin{equation}
    \mathrm{{T^{d,eq}} = {\rm{  }}{T^d}\frac{{{T^{Life}}(u)}}{{{T^{Life}}\left( {{u^{base}}} \right)}}}
    \label{eq:EC7}
\end{equation}
The employed server's expected operating lifetime $\mathrm{T^{Life}}$ under different utilization rates is from \cite{8705686}, which is also taken as the characteristic lifetime $\mathrm{\psi}$. Notably, higher utilization rates lead to increased thermal stress and greater strain on the cooling system due to elevated server operating temperatures, further compounding the failure risk and accelerating the degradation process. 

\subsubsection{Software reliability}
Software failures arise due to latent code defects, malicious cyber intrusions, and resource saturation events, which are random and independent of hardware wear. Similarly, higher utilization will also raise software failure risk due to the increased possibility of resource contention and overload. In \cite{9546936} where the failure probability is modeled as a non-decreasing function, we assume an exponential relationship to represent the non-decreasing nature of server utilization rate. Hence, software reliability $\mathrm{p^S}$ accounts for the impact of deviation in utilization from the baseline $\mathrm{u_0}$ using:
\begin{equation}
    \mathrm{{p^S} = {e^{{\lambda ^S}(u  - {u _0})}}{p^{S,base}}}
    \label{eq:SoftR}
\end{equation}
where, $\mathrm{\lambda^S}$ is the coefficient parameter for software failure risk.

\subsubsection{Server reliability}
Given that hardware and software failures are independent, server reliability $\mathrm{p^{Server}}$ during the dispatched day is dependent on both hardware and software reliability, as follows:
\begin{equation}
    \mathrm{{p^{Server}} ={p^H}( { {{t^{on}} \geq {t^d} + T^{d,eq}} |{t^{on}} \geq {t^d}} )  \cdot {p^S}}
    \label{eq:SeverR}
\end{equation}

\subsection{Reserve Backup Resources}
Based on (\ref{eq:SeverR}), we follow the approach for handling server failures outlined in \cite{9926075}. The number of backup servers $\mathrm{N^B}$ should be sufficient such that the overall reliability exceeds the desired threshold $\mathrm{p^{thr}}$ when the number of failed servers is $\mathrm{N^F}$. If the probability that $\mathrm{N^F}$ servers fail is denoted as $\mathrm{p(N^F)}$, then it can be calculated through binomial distribution where the independent failure probability of each server is considered, as in:
\begin{gather}
    \mathrm{p(N^F \le N^B)  = \sum\nolimits_{{N^F} = 0}^{{N^B}}  {p\left( {N^F} \right)}  \ge {p^{thr}}} \label{eq:ava} \\
    \mathrm{p\left( {{N^F}} \right) = {\begin{pmatrix}
    {\mathrm{N^A}}\\
    {\mathrm{N^F}}
    \end{pmatrix}}{\left( {1 - {p^{Server}}} \right)^{{N^F}}}{\left( {{p^{Server}}} \right)^{{N^A} - {N^F}}}} \label{eq:Bio}
\end{gather}

\section{Linearization of DC Operation Strategy}

A linear and trackable DC operation strategy will assist the solving process. However, the quadratic term and inverse relationship in \eqref{eq:EC6} complicates solving the optimization problem.

First,  the quadratic term $\mathrm{(\sum\nolimits_{t=1}^{24} N_{k,i}^A(t) /24 )^2}$ in \eqref{eq:EC6} is linearized. To simplify the expression, the auxiliary variable $\mathrm{X_{k,i}(t)}$ is introduced to represent $\mathrm{\sum\nolimits_{t=1}^{24} N_{k,i}^A(t) /24 }$, which satisfies constraint \eqref{eq:Xboun}. For notational convenience, the time index $t$ in $\mathrm{X_{k,i}(t)}$ is omitted in the subsequent discussion.
\begin{equation}
    \mathrm{X_{k,i}^{\min } \le {X_{k,i}} \le X_{k,i}^{\max }}
    \label{eq:Xboun}
\end{equation}

Then, another auxiliary variable $\mathrm{Y_{k,i}}$ is introduced to replace the squared term $\mathrm{(X_{k,i})^2}$. To approximate the quadratic relationship between $\mathrm{X_{k,i}}$ and $\mathrm{Y_{k,i}}$, we use the piecewise linearization technique \cite{fourer2002ampl}, where the range of $\mathrm{X_{k,i}}$ is partitioned into $\mathrm{L}$ segments within \{$\mathrm{x^1_{k,i}, x^2_{k,i}, \dots, x^{L+1}_{k,i}}$\} (with $\mathrm{x^1_{k,i} = X^{\min}_{k,i}}$ and $\mathrm{x^{L+1}_{k,i} = X^{\max}_{k,i}}$), and the quadratic function value at each breakpoint $\mathrm{x^{l}_{k,i}}$ is calculated by:
\begin{equation}
    {\mathrm{y^{l}_{k,i} = (x^{l}_{k,i})^2}}, \; \mathrm{\forall l \in \{1, 2, . . . , L + 1\}}
\end{equation}
By defining non-negative weight variables $ \mathrm{\lambda^{l}_{k,i}}$, $\mathrm{X_{k,i}}$ can be expressed as the following interpolation:
\begin{equation}
    \mathrm{X_{k,i} = \sum _{l=1}^{L+1} \lambda^{l}_{k,i} x^{l}_{k,i}},\;\mathrm{\lambda^{l}_{k,i} \geq 0}
    \label{eq:Y1}
\end{equation}
where only two adjacent $\mathrm{\lambda^{l}_{k,i}}$ can be non-zero as per the Special Ordered Set Type 2 (SOS2) constraint for more efficient computation:
\begin{equation}
    \mathrm{\sum_{l=1}^{L+1} \lambda^{l}_{k,i}  = 1} \ni
    {\small
    \begin{cases}
        \mathrm{\lambda^{l}_{k,i} > 0 \quad l\in\{j,j+1\},\:x^{j}_{k,i} \leq X_{k,i} \leq x^{j+1}_{k,i}}\\
        \mathrm{\lambda^{l}_{k,i} = 0 \quad otherwise}
    \end{cases}}
    \label{eq:Y2}
\end{equation}
Moreover, $\mathrm{Y_{k,i}}$ is calculated as:
\begin{equation}
    \mathrm{Y_{k,i} = \sum_{l=1}^{L+1} \lambda^{l}_{k,i} y^{l}_{k,i}}
    \label{eq:Y}
\end{equation}
Based on the above relationship, \eqref{eq:EC6} is transformed into $\mathrm {\sum\nolimits_{k=1}^{K}\frac{C_{k,i}^MY_{k,i}}{T_{k,i}^{PC}  X_{k,i} + T_{k,i}^{FO}  N_{k,i}}}$ with constraints \eqref{eq:Xboun} - \eqref{eq:Y}. However, the inverse relationship still hinders the tractability of the optimization problem. To address this, the auxiliary variable $\mathrm{Z_{k,i} }$ is introduced, which converts \eqref{eq:EC6} into $\mathrm {\sum\nolimits_{k=1}^{K}Z_{k,i}}$. To ensure this equivalence holds, the following constraint is introduced:
\begin{equation}
    \mathrm{{X_{k,i} \cdot Z_{k,i}}T_{k,i}^{PC} + {N_{k,i}}T_{k,i}^{FO}{Z_{k,i}} = C_{k,i}^M{Y_{k,i}}}
    \label{eq:WZY}
\end{equation}
The bounds of $\mathrm{Z_{k,i}}$ are as follows, which are derived by evaluating the fractional expression in \eqref{eq:EC6} at the extreme values of $\mathrm{X_{k,i}}$:
\begin{equation}
    \mathrm{\frac{{{{C_{k,i}^M(X_{k,i}^{\min })}^2}}}{{X_{k,i}^{\min }T_{k,i}^{PC} + {N_{k,i}}T_{k,i}^{FO}}} \le {Z_{k,i}} \le \frac{{{{C_{k,i}^M(X_{k,i}^{\max })}^2}}}{{X_{k,i}^{\max }T_{k,i}^{PC} + {N_{k,i}}T_{k,i}^{FO}}}}
    \label{eq:Zbound}
\end{equation}
To further linearize the fractional term $\mathrm{X_{k,i} \cdot Z_{k,i}}$ in \eqref{eq:WZY}, the McCormick envelope method in \cite{Raghunathan2022Recursive} is employed. The auxiliary variable $\mathrm{Q_{k,i}} $ and following constraints are introduced to make $Q_{k,i}$ approximate the product $\mathrm{X_{k,i}Z_{k,i}}$.
\begin{subequations}
    \begin{align}
        \mathrm{Q_{k,i}} &\geq \mathrm{X_{k,i}^{\min}  Z_{k,i} + X_{k,i}  Z_{k,i}^{\min} - X_{k,i}^{\min}  Z_{k,i}^{\min}} \label{eq:Mc1} \\
        \mathrm{Q_{k,i}} &\geq \mathrm{X_{k,i}^{\max}  Z_{k,i} + X_{k,i}  Z_{k,i}^{\max} - X_{k,i}^{\max}  Z_{k,i}^{\max}} \label{eq:Mc2} \\
        \mathrm{Q_{k,i}} &\leq \mathrm{X_{k,i}^{\max}  Z_{k,i} + X_{k,i}  Z_{k,i}^{\min} - X_{k,i}^{\max}  Z_{k,i}^{\min}} \\
        \mathrm{Q_{k,i}} &\leq \mathrm{X_{k,i}^{\min}  Z_{k,i} + X_{k,i}  Z_{k,i}^{\max} - X_{k,i}^{\min}  Z_{k,i}^{\max}}
    \end{align}
    \label{eq:Mc}
\end{subequations}
In summary, \eqref{eq:EC6} can be reformulated by variables $\mathrm{X_{k,i}}$, $\mathrm{Y_{k,i}}$, $\mathrm{Z_{k,i}}$, $\mathrm{Q_{k,i}}$, and $\mathrm{\lambda_{k,i}}$ as well as constraints \eqref{eq:Xboun}-\eqref{eq:Mc}.

\section{Clustering Model of Heterogeneous Server}

\begin{figure}[t]
    \centering
    \includegraphics[width=0.95\linewidth]{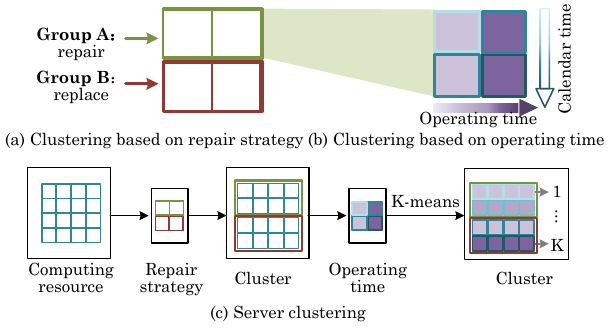}
    \vspace{-9pt}
    \caption{Server clustering framework based on repair strategy and operating time. (a) Servers are grouped according to the expected repair and replacement strategy, since the embodied carbon emissions from the two strategies can differ by up to an order of magnitude. (b) Servers within each group are further clustered using K-means based on their accumulated operating time, enabling the identification of degradation-aware clusters. (c) The complete clustering framework combines both repair strategy and operating time to support carbon- and lifetime-aware server dispatching.}
    \label{Diagram of server cluster}
    \vspace{-9pt}
\end{figure}

To explore the impact of heterogeneous servers on workload migration, this section proposes a clustering framework that integrates repair strategy and operation time, as demonstrated in Fig. \ref{Diagram of server cluster}.

\subsubsection{Grouped based on repair strategy} 
Servers are first classified into \textbf{Group A} and \textbf{Group B} according to the estimated next repair strategy, as shown in Fig.~\ref{Diagram of server cluster}(a). In \textbf{Group A}, server performance remains within the normal operational range, and their cumulative historical maintenance costs $ \mathrm{\sum C^M}$ do not exceed a predefined threshold $ \mathrm{\theta C^R}$. Here, $\mathrm{C^R}$ represents the replacement cost of the entire server. For these servers, the next maintenance cost is modeled as the disk drive replacement cost $\mathrm{C^M}$, reflecting the statistical dominance of disk drives as the most failure-prone components. Conversely, servers exhibiting significant performance degradation or whose cumulative maintenance costs approach the threshold $\mathrm{\theta C^R}$ are grouped into \textbf{Group B}, where the next maintenance cost is predicted to be the full replacement cost $\mathrm{C^R}$.

\subsubsection{Clustering based on operating time} 
As shown in \eqref{eq:EC4}, embodied carbon emissions are highly dependent on the scheduled operating time of the dispatched servers, the servers in \textbf{Group A} and \textbf{Group B} are further clustered in Fig. \ref{Diagram of server cluster}(b). 

We exploit the K-means clustering method \cite{9711558}, which employs past operating time $\mathrm{T^{PO}}$ and past calendar time $\mathrm{T^{PC}}$ as classification criteria. The number of clusters in \textbf{Group A} and \textbf{Group B} is defined as $\mathrm{K^A}$ and $\mathrm{K^B}$, respectively, resulting in a total of $\mathrm{K}$ clusters for each DC. Each cluster consists of $\mathrm{N_{k,i}}$ servers. The centroid values, $\mathrm{T_{k,i}^{PC}}$ and $\mathrm{T_{k,i}^{PO}}$, represent the average past calendar time and past operating time of all servers within cluster $\mathrm{k}$. These centroids and $\mathrm{N_{k,i}}$ are then used in \eqref{eq:EC4} to estimate the embodied carbon emissions.

\section{Numerical Experiments and Analysis}

The performance of the proposed migration strategy is evaluated by analyzing the impact of embodied carbon emission and failure events on the migration strategy, respectively. Moreover, the influence of server utilization rate on carbon emission is also explored. The carbon-aware DC migration strategy is performed using a simulation environment with an Intel Core i5-10210U CPU and 16-GB RAM. YALMIP and GUROBI are employed to solve the MILP model.

\subsection{Parameters Setup}

All case studies have been performed on two interconnected DCs, named DC1 and DC2, as shown in Fig. \ref{Framework}. Each DC is connected to the utility, but is still facilitated with its own carbon-neutral sources, PV and BESS. The utilized carbon emission intensities and PV generation profiles of the DCs are from Arizona and Texas \cite{electricitymaps}. The BESS parameters are listed in Table. \ref{tab:BESS}.
DCs related parameters are listed in Table \ref{tab:Parameters of DCs}. The typical workload profiles of two DCs are demonstrated in Fig. \ref{Embodied carbon emission} (a) \cite{10518095}, where the peak workload occurs during the daytime. We assume that the batch workload accounts for 30 \% of the whole workload, and the remaining 70 \% is interactive. 
\vspace{-20pt}
\begin{table}[h]
    \centering
    \caption{Parameters of BESS}
    \label{tab:BESS}
    \vspace{-6pt}
    \renewcommand{\arraystretch}{1.2} 
    \setlength{\tabcolsep}{8pt} 
    \footnotesize 
    \begin{tabular}{l c l c}
        \hline\hline
        \textbf{Parameter} & \textbf{Value} & \textbf{Parameter} & \textbf{Value} \\
        \hline
        Rated Energy ($\mathrm{kWh}$) & 2000 & Rated Power ($\mathrm{kW}$) & 500 \\
        Efficiency ($\%$) & 95 & Range of SOC ($\%$) & 20--90 \\
        \hline\hline
    \end{tabular}
    \vspace{-12pt}
\end{table}
\begin{table}[h]
    \centering
    \caption{Data Center Parameters}
    \label{tab:Parameters of DCs}
    \vspace{-6pt}
    \renewcommand{\arraystretch}{1.0} 
    \setlength{\tabcolsep}{6pt} 
    \begin{tabular}{l c l c}
        \hline\hline
        \textbf{Parameter} & \textbf{Value} & \textbf{Parameter} & \textbf{Value} \\
        \hline
        $\mathrm{N}$ & 3750 & $\mathrm{\alpha^c}$ & 0.1 \$/kgCO\textsubscript{2} \\
        $\mathrm{\omega^{OC}}$ & 1& $\mathrm{\omega^{EC}}$ & 1\\
        $\mathrm{\omega^{Mig}}$ & 1& $\mathrm{\Delta W_{i \to j,t}^{I,ub}}$ & 10,000 req/s \\ 
        $\mathrm{CI^{EC}}$ (Group A)  &  432 kgCO\textsubscript{2} & $\mathrm{\beta}$  (Group A) & 0.7\\ 
        $\mathrm{CI^{EC}}$ (Group B)  &  4320 kgCO\textsubscript{2} & $\mathrm{\beta}$ (Group B) & 1.3 \\
        $\mathrm{\alpha^{Mig}}$ & $10^{-6}$ cent/req/s & $\mathrm{PUE_{i}}$ & 1.3 \\
        $\mathrm{P^{idle}}$ & 0.1 kW & $\mathrm{P^{peak}}$ & 0.2 kW \\
        $\mathrm{S^{rate}}$ & 20 req/s & $\mathrm{t^{delay,ub}}$ & 0.167 s \\
        $\mathrm{u^{ub}}$ & 0.7 & $\mathrm{ P^{S,base}}$ & 0.99 \\
        \hline\hline\\
    \end{tabular}
    \vspace{-12pt}
\end{table}
\vspace{-12pt}

\subsection{Server Clustering Results}

\begin{figure}[t]
    \centering
    \includegraphics[width=0.8\linewidth]{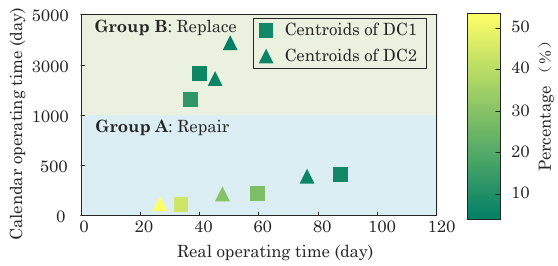}
    \vspace{-9pt}
    \caption{Server clustering results of DC1 and DC2.}
    \label{Distribution of servers}
    \vspace{-9pt}
\end{figure}

Taking DC1 as an example, the calendar operating time and real operating time of the servers are initially generated based on the lognormal distribution and parameters provided in Table III. Servers are subsequently clustered into replace and repair groups according to the estimated maintenance plan. The K-means clustering method is then employed to refine the classification, with the centroids for the repair $\mathrm{K^A}$ and replace $\mathrm{K^B}$ groups set to 3 and 2, respectively -- depicted by the triangles (for DC1) and squares (for DC2) in Fig. \ref{Distribution of servers}. 
\begin{table}[h]
    \centering
    \caption{Log-normal distribution Parameters for DC1 and DC2}
    \vspace{-6pt}
    \label{tab:repair_replace}
    \renewcommand{\arraystretch}{1.0} 
    \setlength{\tabcolsep}{10pt} 
    \footnotesize 
    \begin{tabular}{c *{4}{@{\hspace{0.5ex}}>{\centering\arraybackslash}p{42pt}@{\hspace{0.5ex}}}}
        \hline\hline
        \multirow{2}{*}{\textbf{Parameter}} & \multicolumn{2}{c}{\textbf{Repair}} & \multicolumn{2}{c}{\textbf{Replace}} \\
        \cline{2-3} \cline{4-5}
        & \textbf{DC1} & \textbf{DC2} & \textbf{DC1} & \textbf{DC2} \\
        \hline
        Percentage  & 0.8 & 0.9 & 0.2 & 0.1 \\
        $\mathrm{\mu^{op}}$  & 0.4 & 0.3 & 0.3 & 0.4 \\
        $\mathrm{\sigma^{op}}$  & 0.5 & 0.5 & 0.5 & 0.5 \\
        $\mathrm{\mu^{can}}$  & 1.25 & 1.4 & 1.15 & 1.5 \\
        $\mathrm{\sigma^{can}}$  & 0.3 & 0.3 & 0.3 & 0.3 \\
        \hline\hline
    \end{tabular}
    \vspace{-12pt}
\end{table}

\subsection{Carbon Emission Reduction Results}

\begin{figure}[t]
     \centering
     \includegraphics[width=0.9\linewidth]{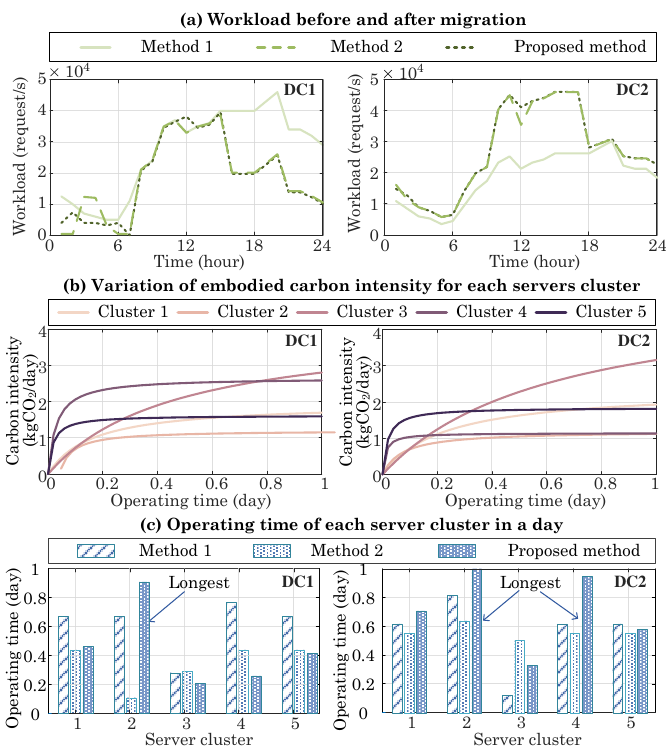}
     \vspace{-9pt}
     \caption{Workload migration strategy and operating strategy of two DCs based on three methods. (a) Workload profiles of DC1 and DC2 before and after migration across the 24-hour scheduling horizon. (b)Embodied carbon intensity of each server cluster as a function of accumulated operating time during a day. Five server clusters are compared in each DC, highlighting their heterogeneity. (c) Total operating time of each server cluster within a day based on the three methods. A  balanced and efficient allocation of workload by prioritizing low-embodied intensity clusters, thereby reducing embodied carbon impact.}
     \label{Embodied carbon emission}
     \vspace{-9pt}
\end{figure}

The proposed method is compared with the following two methods to validate its performance. For fairness, failure events are not incorporated in this section. Following the previous research, the utilization rate is set to its maximum allowable value of 0.7, which corresponds to a maximum allowable delay time of 0.167 seconds.
\begin{itemize}
    \item \textbf{Method 1 (M1)}: No workload migration, considering only operation carbon emission.
    \item \textbf{Method 2 (M2)}: Spatio-temporal workload migration considering only operation carbon emission.
    \item \textbf{Proposed Method}: Spatio-temporal workload migration considering both operation and embodied emissions.
\end{itemize}
As compared to Method I and 2, the proposed method achieves a 21\% reduction in total carbon emissions, which is primarily attributed to two key factors:

\begin{enumerate}
    \item \textbf{Operating Carbon Cost Reduction}: As compared to Method 1, the proposed method and Method 2 reduce operating carbon costs by up to 36\% through spatio-temporal workload migration. Fig. \ref{Embodied carbon emission}(a) illustrates the original workloads and the migrated workloads based on Method 2 and the proposed method. By leveraging higher PV generation during daytime and lower carbon intensity in DC2, both batch workloads of two DCs shift from periods with higher carbon intensity to daytime and interactive workloads are migrated from DC1 to DC2.

    \item \textbf{Embodied Carbon Emissions Reduction}: As compared to Method 1, the proposed method achieves a 6\% reduction in embodied carbon emissions, corresponding to a 6\% decrease in server replacement costs. This improvement stems from the proposed dispatch strategy's explicit consideration of server heterogeneity. As illustrated in Fig. \ref{Embodied carbon emission}(b), the embodied carbon intensities exhibit significant variation across five server clusters and escalate with daily operating time. Consequently, servers with the lowest embodied carbon intensity (e.g., Group 2 in DC1 and Groups 2 and 4 in DC2 -- see Fig. \ref{Embodied carbon emission}(c)) are prioritized for dispatch, leading to extended operating times for these groups. In contrast, the stochastic server dispatch strategy employed in Method 2 results in up to 16\% higher embodied carbon emissions relative to the proposed method. 
\end{enumerate}

\begin{figure}[h]
     \centering
     \includegraphics[width=1\linewidth]{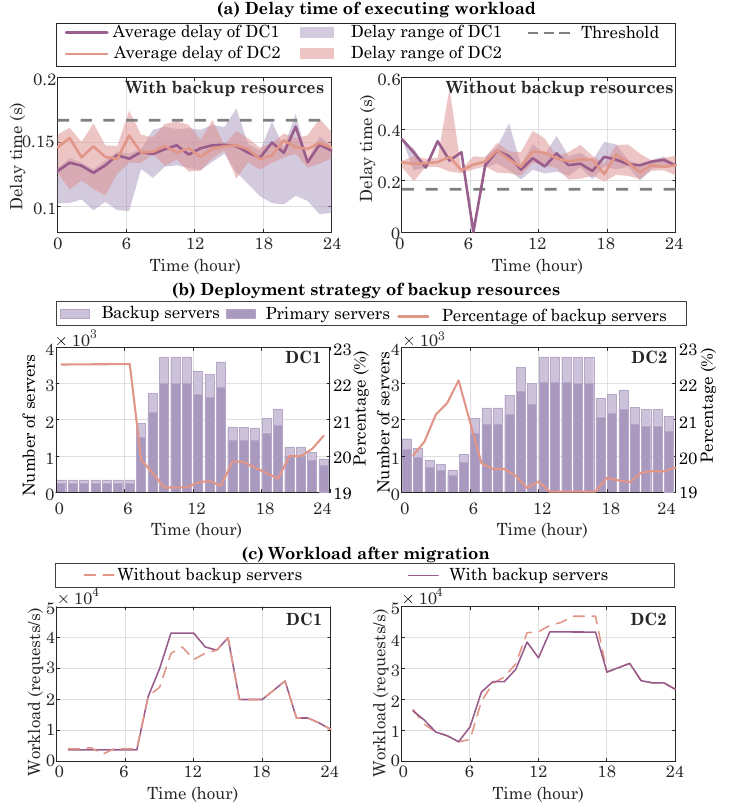}
     \vspace{-9pt}
     \caption{SLA violation and workload profiles with and without backup servers and backup resources deployment strategy: (a) Average and range of workload delay in DC1 and DC2 with and without backup servers. Backup servers significantly reduce peak delays and improve SLA compliance, (b) Deployment of primary and backup servers in both DCs. Backup servers account for a higher proportion during low-demand periods, (c)  Workload profiles after migration. Without backup servers, workloads assigned to DC2 are overestimated, leading to potential overload and SLA violations.}
     \label{SLA violation}
     \vspace{-9pt}
\end{figure}
 
\subsection{Performance Comparison of SLA}
Based on the proposed method, we compared the performance of SLAs under dispatch strategies with and without deploying backup resources for failure events. To evaluate the robustness of the proposed method, we introduce failure events that were stochastically generated based on their failure distribution.

As shown in Fig. \ref{SLA violation}(a), the execution time of workloads consistently exceeds the allowed threshold under the dispatch strategy without backup resources, leading to SLA violations. By deploying backup resources based on the proposed method, SLA violations are significantly reduced to less than 1\%, with the maximum delay time having increased by only 5\%. 

The deployment strategy for backup resources is illustrated in Fig. \ref{SLA violation}(b), where the required backup servers account for approximately 19\% to 23\% of the total dispatched servers. This deployment reduces the workload capacity of data centers, as shown in Fig. \ref{SLA violation}(c). Consequently, the real workload that can be transferred is overestimated in scenarios without backup resources.

\subsection{Impact of server utilization rate}
\begin{figure}[t]
    \centering
    \includegraphics[width=1\linewidth]{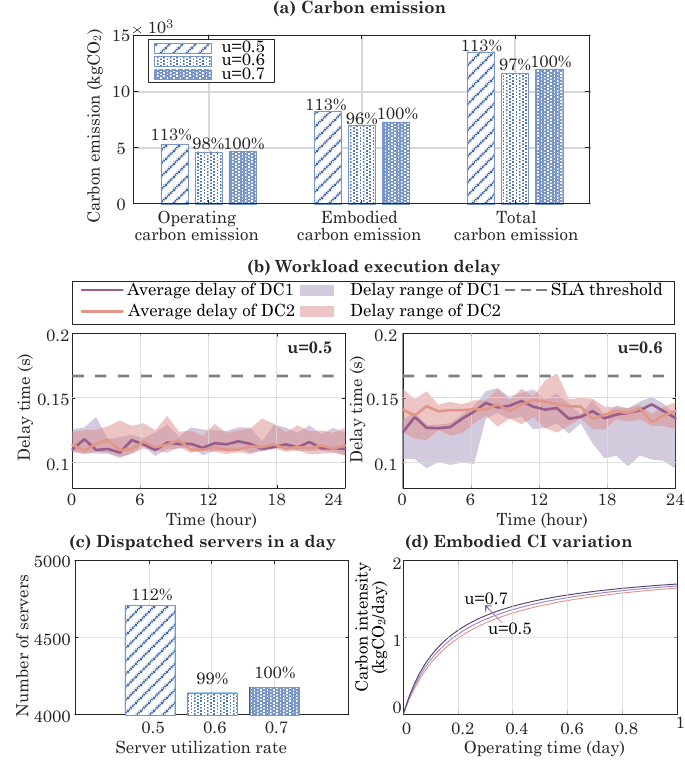}
    \vspace{-9pt}
    \caption{Impact of different server utilization rates on carbon emissions, SLA violations, and server deployment: (a) Operating, embodied, and total carbon emissions under different utilization levels. Moderate utilization reduces total emissions, with 0.6 achieving the lowest. (b)  Workload execution delay across a day under $\mathrm{u = 0.5}$ and $\mathrm{u = 0.6}$. Lower utilization decreases average delays due to under-provisioned resources, (c) Number of dispatched servers required during a day under different utilization rates, (d) Variation of embodied carbon intensity over operating time. Higher utilization leads to higher embodied carbon intensity due to more lifetime degradation.}
    \label{Utilization rates}
    \vspace{-9pt}
\end{figure}

Considering the impact of server utilization rates, we analyze the robustness of the proposed framework under three utilization rates: 0.5, 0.6, and 0.7. As shown in Fig. \ref{Utilization rates}(a), the lowest carbon emissions for DCs are achieved when the server utilization rate $u$ is 0.6. This can be attributed to the minimal number of servers dispatched, as illustrated in Fig. \ref{Utilization rates}(c), which shows the total servers operating for a day. Additionally, while the embodied carbon intensity decreases at lower utilization rates, as depicted in Fig. \ref{Utilization rates}(d), this reduction cannot offset the increased embodied carbon emissions resulting from the longer lifetime consumption due to more dispatched servers. In terms of SLA violation conditions, as shown in Fig. \ref{Utilization rates}(b) and Fig. \ref{SLA violation}(a), the best performance is achieved when $u$ is 0.5 due to the over-deployment of servers. Although a few SLA violations occur when $u$ is 0.6, which is less than the scenario where $u$ is 0.7.

\begin{figure}[t]
    \centering
    \includegraphics[width=1\linewidth]{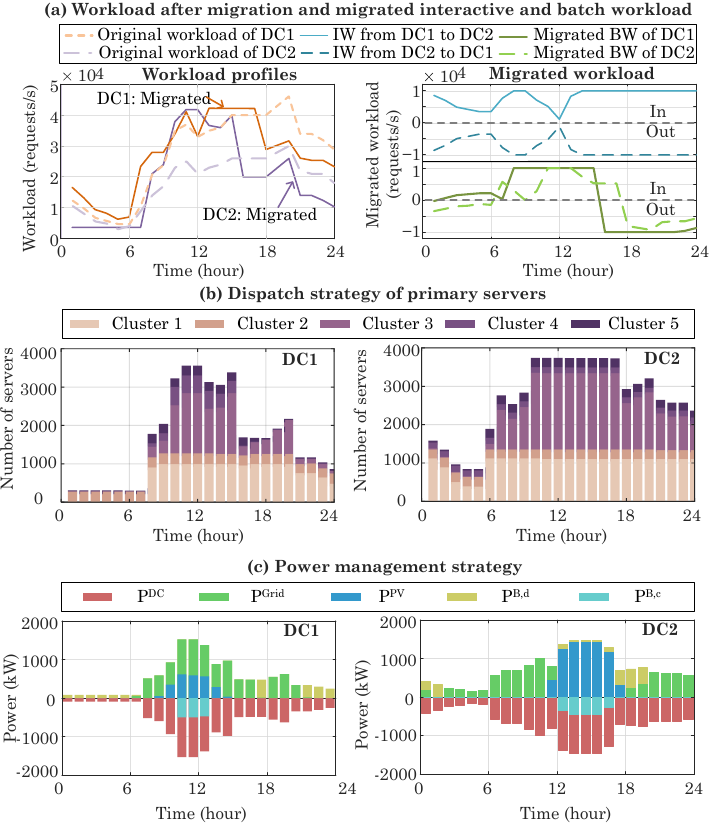}
    \vspace{-15pt}
    \caption{Operating strategy of DCs when $\mathrm{u}$ is 0.6. (a) Workload after migration and direction of migrated interactive and batch workloads. DC2 handles more interactive workloads due to lower carbon intensity. (b) Dispatch profiles of primary servers by cluster. (c) Power management strategy of DCs, where $\mathrm{P^{DC}}$, $\mathrm{P^{Grid}}$, $\mathrm{P^{PV}}$, $\mathrm{P^{B,d}}$ and $\mathrm{P^{B,c}}$ denote the total power demand of DC, external grid, locally generated photovoltaic power, battery discharge and charge power, respectively. }
    \label{Operating strategy at $u$ is 0.6.}
    \vspace{-9pt}
\end{figure}

Taking into account both carbon emissions and SLA conditions, the best fit of the server utilization rate is obtained at a value of 0.6. Considering this as an optimal utilization rate, the workload migration strategy, computing resource dispatch strategy, and power management strategy are illustrated in Fig. \ref{Operating strategy at $u$ is 0.6.}, respectively. Fig. \ref{Operating strategy at $u$ is 0.6.}(a) shows that the peak workload in DC1 is due to batch workloads being mostly allocated to daytime with lower carbon intensity and abundant PV generation, while the peak workload in DC2 not only attribute to migrated batch workload, but also the interactive workloads from DC1. Fig. \ref{Operating strategy at $u$ is 0.6.}(b) indicates that server clusters with lower aging levels are more actively utilized, while higher aging servers are usually utilized during high-demand periods. The dispatch strategy demonstrates the effectiveness of the proposed degradation-aware and reliability-constrained clustering framework. In Fig. \ref{Operating strategy at $u$ is 0.6.}(c), the power management strategy reveals: excess PV generation enables battery charging and reduces reliance on the grid, while in the morning and evening peak hours, battery discharging complements grid supply to reduce carbon emission. This coordinated power scheduling highlights the role of hybrid energy sources in minimizing carbon emissions.
   
\section{Conclusion}

This paper proposes a comprehensive carbon- and reliability-aware optimization framework for spatio-temporal workload migration in distributed DCs. A server lifetime-oriented dispatch strategy is proposed to account for server heterogeneity, aiming to minimize both operational and embodied carbon emissions. Furthermore, a chance-constrained backup resource allocation model is developed to ensure SLA compliance under uncertainty. Numerical analysis conducted on two interconnected DCs demonstrates that: 
\begin{enumerate}
    \item Compared to the baseline method, the proposed strategy achieves up to 21\% reduction in total carbon emissions, owing to the prioritized use of low-embodied-emission server clusters.
    \item By incorporating backup resource allocation, the framework significantly improves service reliability, with SLA violations reduced to less than 1\%.
    \item A sensitivity analysis on server utilization rates reveals that a moderate utilization level (e.g., 0.6) provides an optimal balance between energy efficiency and operational reliability of DCs. 
\end{enumerate}
These results validate the effectiveness and practicality behind the proposed framework in enabling sustainable and reliable operation of distributed DCs.

\bibliographystyle{IEEEtran}
\bibliography{ref}

\end{document}